\documentclass[aps,12pt]{revtex4}
\usepackage[dvips]{epsfig}
\usepackage{amsmath,amssymb}

\parskip=\medskipamount

\newcommand{\eq}[1]{(\ref{#1})}
\newcommand{\fig}[1]{Fig.\ref{#1}}

\newcommand{\be}{\begin{equation}}
\newcommand{\ee}{\end{equation}}

\newcommand{\barr}{\begin{array}}
\newcommand{\earr}{\end{array}}

\newcommand{\beqn}{\begin{eqnarray}}
\newcommand{\eeqn}{\end{eqnarray}}

\newcommand{\bs}{\begin{subequations}}
\newcommand{\es}{\end{subequations}}

\newcommand{\bw}{\begin{widetext}}
\newcommand{\ew}{\end{widetext}}

\newcommand{\mbf}{\mathbf}
\newcommand{\mcl}{\mathcal}

\newcommand\disp{\displaystyle}

\begin{document}

\title{Application of a two--length scale field theory to the solvation
of charged molecules: I. Hydrophobic effect revisited}
\author{G.Sitnikov$^{1,2}$, S.Nechaev$^{3,4}$, M.Taran$^2$, A.Muryshev$^2$}
\affiliation{$^1$Moscow Institute of Physics and Technology, Institutskaya str.9,
141700, Dolgoprudnyi, Russia \\ $^2$Algodign LLC, Bolshaya Sadovaya str., 8(1),
123379, Moscow, Russia \\ $^{3}$LPTMS, Universit\'e Paris Sud, 91405 Orsay Cedex,
France \\ $^{4}$L.D.Landau Institute for Theoretical Physics, 117334 Moscow, Russia}

\date{May 12, 2005}

\begin{abstract}
On a basis of a two--length scale description of hydrophobic interactions we develop
a continuous self--consistent theory of solute--water interactions which allows to
determine a hydrophobic layer of a solute molecules of any geometry with explicit
account of solvent structure described by its correlation function. We compute the
mean solvent density profile $n(r)$ surrounding the spherical solute molecule as
well as its solvation free energy $\Delta G(r)$. We compare the two--length scale
theory to the numerical data of Monte--Carlo simulations found in the literature and
discuss the possibility of a self--consistent adjustment of the free parameters of
the theory. In the frameworks of the discussed approach we compute also the
solvation free energies of alkane molecules and the free energy of interaction
$\Delta G(D)$ of two spheres of radius $R$ separated by the distance $D$. We
describe the general setting of the self--consistent account of electrostatic
interactions in the frameworks of the model where the water is considered not as a
continuous media, but as a gas of dipoles. We analyze the limiting cases where the
proposed theory coincides with the electrostatics of a continuous media.
\end{abstract}

\maketitle


\section{Introduction}
\label{intr}

We consider the current work as a step towards the development of a continuous
thermodynamic theory of solute--solvent interactions. The construction of such a
kind of a theory is of extreme importance for biochemical applications connected
with the reliable and fast determination of the protein--ligand binding constants in
solvent (water) with the precision comparable to the accuracy of explicit
molecular--dynamical simulations.

There are few kinds of approaches which take into account the effect of fluctuating
media on interactions between solvated molecules. Tentatively we can split them into
the following three groups:

i) "Explicit approaches". Generally such methods are based on numerical simulations
taking into account motion of many water molecules. Such approaches are very time
consuming because they demand huge computational resources.

ii) "Implicit approaches". These: a) either exploit the heuristic concepts of the
"surface area accessible by the solvent" and of the "width of the hydrophobic
layer", or b) make some effective renormalization of "bare" (i.e. vacuum) parameters
of interacting molecules: in the frameworks of these approaches the effect of water
on thermodynamic properties of dissolved substances is computed on the basis of some
additional data on solubility, evaporation etc---see, for example,
\cite{george,gav,finkelstein}.

iii) "Intermediate approaches". The works belonging to that group (including our
one), offer attempts which, on one hand, get rid of explicit numerical simulation of
water molecules, and, on the other hand, preserve an effective account for water.
The development of these methods on the basis of a clear understanding of a physical
origin of solute--water interactions could, at least, give a better control for the
parameters used in implicit approaches.

Most commonly, in the "implicit approaches" it is assumed \cite{surf,priv} that the
free energy of solvation is proportional to the solvent accessible surface area of a
solute molecule. It is obviously true for solutes much larger than a water molecule.
However, for the solutes of sizes comparable to the size of water molecules, as well
as for the solutes of complex geometry, the question of the determination of the
corresponding hydrophobic layer, being of great importance, is often resolved by
means of purely empiric conjectures. For example, it is supposed usually that the
hydrophobic layer is equal to the surface covered by a center of a water molecule
"rolling" around the solute surface. Some modifications of the method deal with the
so-called "molecular surface area" -- see, for example, \cite{msa}. Another method
of implicit account for the solvent in the so-called "Gaussian approximation" has
been proposed by \cite{karp}.

Among the theoretical attempts to develop a constructive theory of solvent--solute
interactions, the following groups of works (to our point of view) deserve special
attention:

1. The detailed computation of the solvent density--density correlation function in
the presence of dissolved substances (with or without electrostatics): a) either
introducing weighted densities and explicitly designing the form of the density
functional which is entirely dictated by physical constraints \cite{rosen}, or b) by
the construction of an appropriate "bridge functional" (i.e. closure relation) for
the correlation function satisfying the hierarchy of integral equations
\cite{3drism}. On the basis of the correlation function all the thermodynamic
quantities can be easily determined.

2. The computation of the averaged profile of the solvent density described by the
fluctuating field in presence of solute molecules (with or without electrostatic
interactions). The solvent structure on small scales is taken into account by its
bulk correlation function, while the effect of solute solvation is considered by
forcing the total solvent density to be zero inside the solute molecule (see, for
example, \cite{chand4} and references therein). Minimization of the corresponding
free energy allows one to compute the density and the solvation free energy.

3. The "Scaled Particle Theory"(SPT) \cite{spt,spt1} and the "Information Theory"
(IT), developed in the series of papers \cite{pratt}. Both these approaches, SPT and
IT, exploit the relation between the probability to find a cavity of a given shape
containing some number of particles, with the solvation free energy of a solute
molecule of that shape. The probability of a cavity formation in an ensemble of
fluctuating particles (water molecules) is computed on the basis of semi--empirical
combinatorial arguments.

Our solvation model borrows its ideas in the second group from the list above,
namely from the series of works \cite{chand1,chand2,chand3} where a two--length
scale description of hydrophobic interactions was developed. In this approach the
solvent density is decomposed in two components, the slowly varying field describing
the mean solvent (water) density, and the short--ranged density fluctuations. We
believe that such an approach is optimal from different points of view: on one hand
it is physically clear, being "semi-microscopic", and on other hand it can be used
as a constructive computational tool of account for water, much faster than the
corresponding explicit approaches but without the loss of the precision. So, our aim
is to construct a theory of solute--water interactions which would allow to
determine the hydrophobic layer of solute molecules of any geometry with account of
water structure described by its correlation function. The parameters of such a
theory will be determined by testing them for solutes with simple geometries and
comparing these cases with the experimental data. Briefly, the main ingredients of
our work are as follows:

i) We consider water as a continuous inhomogeneous media and describe it by a
fluctuating continuous density field; the discreteness of the water structure on
small scales is taken into account through the fluctuations controlled by the water
correlation function $\chi$ in the bulk;

ii) The solvation free energy of the substance situated in water is obtained in the
self--consistent mean--field approximation based on slightly modified
Ginzburg--Landau--Chandler functional \cite{chand1,chand2}.

Our work is devoted to the calculation of the solvation free energy and of the free
energy of interactions of solute molecules of any arbitrary shapes.

Specifically, we start with a two--scale Ginzburg--Landau Hamiltonian and perform an
analytical integration over the short--ranged fluctuations. This provides a
single--variable Hamiltonian yielding an integro--differential equation. The
solution of this equation ensures the equilibrium mean density profile of the water,
and the corresponding value of the Hamiltonian gives the solvation free energy. We
pay a special attention to the physical clarity and consistency of all steps of
derivation of our model. We discuss the hidden dangers on the way and compare our
approach to the Monte--Carlo numerical simulations carried out in \cite{chand4}. We
determine in a self--consistent way the width and the geometry of a hydrophobic
layer surrounding the solute molecules. So, we believe that a very important
auxiliary goal of our work concerns the prediction of the relevant number of water
shells (the "solvent accessible surface area") around a solute molecule.

Summarizing the said above, the advantages of the fluctuational approach with
respect to other phenomenological theories of "implicit water" are as follows:

a) There is no difference in computations of the solvation free energy and of the
free energy of interactions of solute molecules;

b) The detailed structure of the solvent is taken into account on small scales;

c) We do not need any special determination of the "hydrophobic layer" by "rolling"
the water molecule around the solute surface. The width and the geometry of a
hydrophobic layer is computed on the basis of: a) the true geometry of a solute
molecule, and b) the supposition about the solute--selvent and solvent--solvent
interactions.

d) The theory contains small number of adjustable parameters, which in turn can be
determined from the independent experiments (either numerical, or real).

Our paper is organized as follows. In the Section II we describe the model and
derive the basic equations for the equilibrium and the full densities, and for the
solvation free energy, as well as we solve numerically the obtained equations,
compare them to the numerical Monte--Carlo simulations of the work \cite{chand4} and
propose the way of tuning the adjustable parameters of the Hamiltonian; in the
Secion III we compute the solvation free energy of alkanes, as well as the free
energy of interaction of two spherical solute molecules; the results of the work are
summarized in the Section IV, where we also derive the equations for electrostatics
in fluctuating dipolar environment.

\section{The model}
\label{model}

In this Section we present the self--consistent description of solvation which takes
into account simultaneously: i) the smoothly varying average density field $n{(\bf
r})$, and ii) the water structure on small scales described by the density--density
correlation function $\chi({\bf r},{\bf r}')$.

Following the general scheme of the works \cite{chand3}, we describe the solvent by
two fields: $n({\bf r})$ -- the average density varying on long distances, and
$\omega({\bf r})$ -- the small-scale fluctuating field controlled by the correlation
function of the water $\chi({\bf r},{\bf r}') =\left<\omega({\bf r}) \omega({\bf
r}')\right>$ in the bulk. The solvent cannot penetrate into the volume occupied by
the solute molecule hence the total solvent density $\rho({\bf r})=n({\bf
r})+\omega({\bf r})$ is nullified inside the solute: $\rho({\bf r})=0$ for all ${\bf
r}\in v_{\rm in}$, where $v_{\rm in}$ is the volume occupied by the solute molecule.
The fact that we force the total solvent density $\rho({\bf r})=n({\bf
r})+\omega({\bf r})$ to be zero inside the solute, results in an effective
interactions between $n({\bf r})$ and $\omega({\bf r})$. In general, one can permit
also the direct coupling between $n({\bf r})$ and $\omega({\bf r})$ everywhere in
the solute.

The thermodynamic properties of the solvent (water) in the bulk are obtained using
the self--consistent Ginzburg--Landau (GL) free energy functional which describes
the liquid--vapor phase transition. The average solvent density $n({\bf r})$ plays
the role of an order parameter. The interactions between long-- and small--scale
fields $n({\bf r})$ and $\omega({\bf r})$ can enforce the "drying", shifting the
liquid--vapor transition. Minimizing the corresponding free energy functional, we
compute the desired density profile and the solvation free energy.

It should be noted that the fully microscopic description of the solute--solvent
interactions certainly does not require introducing of any scale
separation---everything is described by the single microscopic field $\rho({\bf
r})$. However in such a description the interactions between water molecules should
also be described on the microscopic level. Such kind of theory has been developed
in the works \cite{rosen}. It seems to us that due to the heavy machinery, the
application of such approach to the solvation and interactions of objects of complex
architecture in water is yet rather restrictive for practical purposes. So, the
nature of two different scales in the approach chosen in our work, is the
consequence of the mean--field (GL) description of the water--water interactions.

Before going into details let us point out what is different in our consideration
with respect to the original two--length scale theory of Chandler et al.
\cite{chand1,chand2,chand3,chand4}.

i) We do not divide the space in "cells" for the large--scale field $n({\bf r})$ (as
in \cite{chand4}). We describe the system by two fully continuous densities. The
equilibrium density is determined selfconsistently without additional averaging  over
small--scale fluctuations. We believe that this would result in better description of
solvation of objects of rough shape and of the free energy of interactions between
solutes located close to each other.

ii) We can tune the potential $W(n({\bf r}))$. Currently $W(n({\bf r}))$ describes
the behavior of the system exactly at the vapor—-liquid phase transition, but our
consideration can be straightforwardly generalized to other situations.

iii) We can consider the charged objects in the frameworks of the "discrete
electrostatics", where the density $n({\bf r})$ becomes the density of water
dipoles. The set of basic equations for the density field $n({\bf r})$ and the
electric field $\phi({\bf r})$ is obtained in Section III, while in the forthcoming
paper \cite{sineta} we numerically solve the derived system of equations.

iv) We can incorporate in the theory the direct Van-der-Waals solute—-solvent
potentials and consider the smooth solute--solvent boundary (see the Section
\ref{disc} for more details).

\subsection{The Hamiltonian}

We describe the solvent in absence of any solute molecules by the following
two--length scale Hamiltonian
\be \mcl{H}_0[\omega,n]=\frac{1}{2}\int \omega({\bf r})
\chi^{-1}({\bf r},{\bf r}') \omega({\bf r}') \, d{\bf r}d{\bf r}' + \int
\left\{\frac{a}{2} \left(\nabla n({\bf r})\right)^2+ W(n({\bf r}))\right\}d{\bf r}
\label{eq:PSH}
\ee
where $n({\bf r})$ is the smoothly changing (average) solvent density (also
considered as the "order parameter" of the theory); $\omega({\bf r})$ is the field
corresponding to the short--ranged density fluctuations; $\chi(|{\bf r}-{\bf r}'|)$
is the solvent correlation function in the bulk; $a$ is the phenomenological
parameter which requires the microscopic determination---see the discussions below;
and the self--consistent potential $W(n({\bf r}))$ is chosen in the common form of
the standard Ginzburg--Landau (GL) expansion for the order parameter $n({\bf r})$ as
the fourth--order polynomial:
\be
W(n)= \frac{b}{2} (n-n_1)^2\,(n-n_2)^2 \qquad (0\le n_1\le n_2\le 1) \label{eq:GLP}
\ee
where $n_1$ and $n_2$ are the values of the order parameter $n$ in the vapor and
liquid phases correspondingly (below we set, if not specified, $n_1=0$) and $b$ is
the coupling constant which in combination with the parameter $a$ defines the
surface tension (see \eq{eq:solv_fr_en} below).

The Hamiltonian \eq{eq:PSH} consists of two (still decoupled) parts. The first term
describes the non-local small--scale Gaussian fluctuations of the solvent. These
fluctuations extend to the distances controlled by the solvent density correlation
function in the bulk $\chi(|{\bf r}-{\bf r}'|)$,
\be
\chi(|{\bf r}-{\bf r}'|)=\langle \omega({\bf r}) \omega({\bf r}')\rangle
\ee
The second term describes the fluctuations of the self--consistent potential
\eq{eq:GLP} in the system which could exhibit a phase transition. In our case this
is a liquid--vapor (i.e. "drying") phase transition.

\begin{figure}[ht]
\epsfig{file=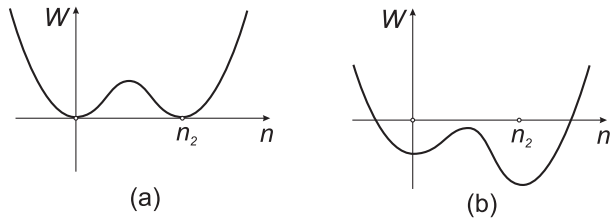,width=8cm} \caption{Sample plots of various shapes of the
potential $W(n)$. The case (a) corresponds to \eq{eq:GLP} and is analyzed in this
paper.} \label{fig:demo}
\end{figure}

The choice of a functional form of the potential \eq{eq:GLP} is rather arbitrary
being to some extent a question of a taste. However it should obey some basic
requirements which are imposed mainly by physical reasons. As it has been mentioned
already, the Hamiltonian \eq{eq:PSH} with the potential \eq{eq:GLP} should describe
the vapor--liquid (i.e. "drying") phase transitions. This should be valid only in
the vicinity of the solvated object, thus manifesting the nature of a hydrophobic
effect. The presence of a drying transition near the solute surface shifts
effectively the parameters of the bulk media to the region of the liquid--vapor
coexistence \cite{chand2}. The general form of the Hamiltonian describing the
liquid--water coexistence is well known -- it is the two--well potential usually
parameterized by the GL expansion---see \fig{fig:demo}. The parameterization being
rather arbitrary, might be taken not only in the polynomial form as in the GL
theory, but also in any form that correctly describes the region of phase
coexistence under conditions close to normal. We believe that the GL expansion is
well enough to reflect the major features of the physical picture. The choice of a
more sophisticated form of the potential \eq{eq:GLP} entering in the large--scale
part of the Hamiltonian \eq{eq:PSH} is the possible way of the refinement of the
model in future within the frameworks of the mean--field approach.

In \fig{fig:demo} for generality we show broader class of potentials $W(n)$ than
described by Eq.\eq{eq:GLP}, allowing the nonsymmetric shapes (see \fig{fig:demo}b).
The profiles in \fig{fig:demo}a,b are drawn for $n_1=0$.

The potential $W(n)$ in the GL expansion depends upon four parameters ($a,\, b,\,
n_1,\, n_2$), two of them (the densities $n_1$ and $n_2$) are fixed in each phase.
The parameters $a$ and $b$ are considered as the free adjustable coefficients of the
theory. Supposing slightly more general form of the potential \ref{eq:GLP},
$\tilde{W}(n)= W(n) + \mu (n-n_1)$, shown in \fig{fig:demo}b, we may consider the
relative difference (RD) between the energy densities in each phase, i.e. the
difference between the values at the minima of $\tilde{W}(n)$ (see \fig{fig:demo}b)
as an extra adjustable parameter. It is easy to see that this difference is closely
connected to the asymptotic behavior of the solvation energy at large solute sizes.
Actually, the RD contributes to the volume--dependent part of the solvation energy.
It is obvious that the term which scales linearly with the volume, tends to
$\frac{1}{3}P V_{\text{solute}}$ as the solute volume $V_{\text{solute}}$ increases
\cite{spt} (here $P$ is the liquid pressure). However the simple quantitative
estimation shows that on the nanometer scale the volume contribution is negligible
in comparison with the surface contribution to the solvation energy \cite{chand4}.
Thus we may neglect the difference in minima of the potential shown in
\fig{fig:demo}b and consider everywhere further the potential symmetric upon the
phase change---see \fig{fig:demo}a. Just such functional form of the potential is
proposed above in \eq{eq:GLP}.

We can find the equilibrium densities minimizing the Hamiltonian (\ref{eq:PSH}) with
respect to the (still decoupled) fields $n({\bf r})$ and $\rho({\bf r})$:
\bs \beqn
\int\chi({\bf r}-{\bf r}')\omega({\bf r}')\,d{\bf r}'=0  \label{eq:ED1} \\
-a \Delta n({\bf r})+\frac{\delta W(n({\bf r}))}{\delta n({\bf r})}=0
\label{eq:ED2}
\eeqn \es
In absence of any interactions between the fields $\omega$ and $n$, the solution of
(\ref{eq:ED1}) postulates the homogeneity of the liquid $\omega(\mbf{r})=
\text{const}$, while Eq.\eq{eq:ED2} describes: a) the plain vapor--liquid interface
supplied with the boundary conditions $n(\mbf{-\infty})\equiv n_1=0$
($n(\mbf{\infty})= n_2$) in the one--dimensional case, and b) the density profile
near the surface of extended macroscopic object in the two-- and three--dimensional
case with boundary conditions $n=0$ at the solute surface and $\mbf{\nabla}n=0$ for
$|\mbf{r}|\to \infty$. Everywhere further in this Section for simplicity we consider
the spherical objects only (the solvation of objects of other shapes is discussed at
lengths of the Section III).

Qualitatively these solutions can be analyzed by computing the first integral of
motion (as in the classical mechanics \cite{ll_1}). Writing the first term in
Eq.\eq{eq:ED2} in 3D as $\Delta n(r) = n''(r)+\frac{2}{r}n'(r)$, multiplying
\eq{eq:ED2} by $n'(r)\equiv \frac{dn(r)}{dr}$ and integrating it from $r_{\rm in}$
(the radius of the sphere) to $\infty$, we obtain:
\be -a\int_{r_{\rm in}}^{\infty} \frac{d}{dr}\left[\frac{(n'(r))^2}{2}\right] dr
- a \int_{r_{\rm in}}^{\infty} \frac{2}{r} (n'(r))^2 dr + \int_{r_{\rm in}}^{\infty}
\frac{dW(r)}{dr} dr= 0; \qquad n'(\infty)=0 \label{eq:integr}
\ee
Hence, we get from \eq{eq:integr} the following relation:
\be
\label{mech_analog}(n'(r))^2\Big|_{r=r_{\rm in}}=\int_{r_{\rm in}}^{\infty}
\frac{4}{r}(n'(r))^2 dr+ \frac{2}{a}[W(n(r_{\rm in}))-W(n(\infty))]
\ee
Having the equation \eq{mech_analog} and the exact functional form of the potential
\eq{eq:GLP} in hand, we know the value of $n'(r)$ at the point $r=r_{\rm in}$ which
determines the qualitative behavior of the density profile. In our case the second
integrand vanishes, hence the density monotonically increases from $n_1$ at the
point $\mbf{r}_{\text{in}}$ to $n_2$ for $|\mbf{r}|=\infty$. In practice, the liquid
density reaches quickly the bulk value at some distance outside the solute surface.
The analysis of an asymmetric potentials with RD can be performed within the same
formalism.

The sense of the parameters $a$ and $b$ becomes transparent in the limit of large
spherical solute molecules $r_{\rm in}\gg \delta$, where $\delta$ is the
characteristic size of the liquid--vapor interface (see below). Substituting
\eq{eq:GLP} into \eq{eq:ED2} and replacing the interaction of the field $n({\bf r})$
with the solute molecule by the boundary condition, we get
\be \left\{\begin{array}{l}
-a\Delta n({\bf r})+ 2 b (n({\bf r})-n_{\text{avr}})(n({\bf r})-n_1)(n({\bf
r})-n_2)=0 \medskip \\ n({\bf r}=r_{\rm in})=n_1
\end{array}\right. \label{eq:surf}
\ee
where $\disp {n_{\text{avr}}}=\frac{n_2+n_1}{2}$. Supposing in \eq{eq:surf} that
$r\gg \delta$, we can expand the 3D Laplacian
$$
\Delta = \left.\frac{d^2}{dr^2}+\frac{2}{r}\frac{d}{dr}\right|_{r\gg \delta} =
\frac{d^2}{dr^2} + O\left(\frac{\delta}{r}\right)
$$
and obtain the effective 1D equation
\be \left\{\begin{array}{l}
\displaystyle-a\,\tilde{n}''(r) + 2 b\, (\tilde{n}(r)-n_{\text{avr}})
(\tilde{n}(r)-n_1)(\tilde{n}(r)-n_2)=0 \medskip \\ \tilde{n}(r_{\rm
in})=n_{\text{avr}}
\end{array} \right. \label{eq:surf1}
\ee whose solution reads
\be
\tilde{n}(r) = \frac{n_2-n_1}{2} \tanh\left(\frac{r-r_{\rm in}}{\delta}\right) +
\frac{n_2+n_1}{2}; \qquad \delta = \frac{2}{n_2-n_1}\sqrt{\frac{a}{b}}
\label{eq:surf_sol}
\ee
It is clear that the symmetric potential of GL form leads to the one--parameter
theory. This parameter $\delta$ fixes the width of the phase separation interface.
The solvation free energy $\tilde{G}$ per the spherical solute area $4\pi r_{\rm
in}^2$ in the limit of large molecules can now be expressed as follows
\be
\barr{lll} \disp \frac{\tilde{G}}{4 \pi r_{\rm in}^2} & = & \disp \frac{1}{4\pi
r_{\rm in}^2} \int_{-\delta}^{\delta} \left\{\frac{a}{2}(\nabla \tilde{n})^2+
\frac{b}{2}(\tilde{n}-n_1)^2(\tilde{n}-n_2)^2\right\} 4\pi r^2 dr
\medskip \\ & = & \disp \frac{\sqrt{a b}}{6}(n_2-n_1)^3 +
O\left(\frac{\delta}{r_{\rm in}}\right) \earr \label{eq:solv_fr_en}
\ee
where the coefficient
$$
\sigma =\displaystyle \frac{\sqrt{a b}}{6}(n_2-n_1)^3
$$
has a sense of a surface tension. Note that for $0<\frac{\delta}{r_{\rm in}} \ll 1$
in the leading order of the expansion \eq{eq:solv_fr_en} the true boundary condition
$n(r_{\rm in}-\delta)=0$ cannot be distinguished from the approximate one, $n(r_{\rm
in})=n_{\text{avr}}$.

\subsection{The correlation function}

The important component of our model is the bulk correlation function, $\chi(|{\bf
r}-{\bf r}'|)$, of a solvent. It appears as an input into the theory. Recall some
basic facts concerning the function $\chi(|{\bf r}-{\bf r}'|)$. We have by
definition:
\be
\chi(|{\bf r}-{\bf r}'|)=\overline{n}\delta({\bf r}-{\bf r}')+\overline{n}^2 h(|{\bf
r}-{\bf r}'|) \label{corrfun}
\ee
where $\overline{n}$ is the mean solvent density in the bulk, $h({\bf r}-{\bf
r}')=g({\bf r}-{\bf r}')-1$ and the function $g(|{\bf r}-{\bf r}'|)$ has very clear
physical meaning:
\be
n^2({\bf r}) g({\bf r})=V^{-1}\big{\langle} \sum_{i\neq j}\delta({\bf r} - {\bf
r}_i+{\bf r}_j)\big{\rangle}
\ee
The function $g({\bf r})$ may be determined from experimental data, or numerical
simulations, or, say, from some selfconsistent integral equation with corresponding
bridge functional \cite{sarkisov}.

The following relation is the direct consequence of the definition of $\chi({\bf
r}-{\bf r}')$:
\be
\overline{n} \int \Big(g(|{\bf r} - {\bf r}'|)-1\Big)\, d{\bf r}'=4\pi \overline{n}
\int h(r) r^2 dr=-1
\ee
The inverse correlation function $\chi_{\rm in}^{-1}({\bf r},{\bf r}')$ is defined
as follows
\be
\left\{\barr{rcll} \disp \int_{v_{\rm in}}\, d{\bf r}''\chi_{\rm in}^{-1}({\bf
r},{\bf r}'') \chi(|{\bf r}''-{\bf r}'|) & = & \delta({\bf r}'-{\bf r}) & \qquad
\mbox{for ${\bf r}\in v_{\rm in}$} \medskip \\ \disp \chi^{-1}_{\rm in}({\bf r},{\bf
r}'') & = &  0 & \qquad \mbox{for ${\bf r}\in v_{\rm out}$} \earr \right.
\label{chi_inv}
\ee
Let us pay attention that the integration in \eq{chi_inv} is spread only to the
volume $v_{\rm in}$ occupied by the solute and hence the function $\chi_{\rm
in}^{-1}({\bf r},{\bf r}')$ is not translationally invariant.

\begin{figure}[ht]
\epsfig{file=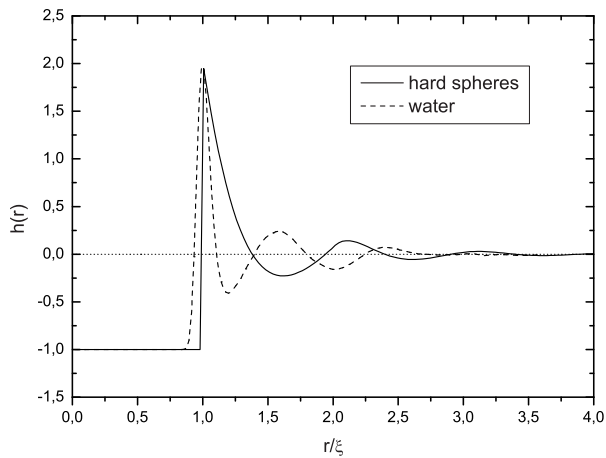,width=9cm} \caption{Correlation function of water {\it vs}
Percus--Yevic correlation function of hard spheres. The distance is measured in
dimensionless units $r/\xi$ ($\xi=2.78$ \AA), the dimensionless bulk density is
$n_2=0.7$.} \label{fig:8}
\end{figure}

We have used in our work two different correlation functions: i) the correlation
function of water obtained from the numerical molecular--dynamical simulations
\cite{corr_water}, and ii) the correlation function of hard spheres constructed on
the basis of solution of self--consistent integral equation completed by the
Percus-Yevic bridge functional \cite{py,PYanalyt}. The functions $h(|{\bf r}-{\bf
r}'|)=g(|{\bf r}-{\bf r}'|)-1$ of water and of hard sphere liquid are shown in
\fig{fig:8}. Both functions correspond to the dimensionless bulk density $n_2=0.7$.
Some approximate equations of the theory of liquids in the statistical
thermodynamics of classical liquid systems one can find, for instance, in the review
article \cite{sarkisov}.

Working simultaneously with two different correlation functions (of water and of
hard spheres) we able able to check the sensitivity of the physical observable
quantities (as the density and the free energy) to the details of the correlation
function, as well as to verify the convergence of our numerical scheme: far away
from the solute surface the results of computations should not be sensitive to the
particular choice of the correlation function.

The correlation function defines the natural scale, $\xi$, in the theory. In what
follows we fix $\xi=2.78$ \AA~ as the distance till the first peak in the
correlation function of water.

\subsection{The partition function}

Now we turn to the solute--solvent interactions. To take into account the effect of
the solute solvation, we demand the total density $\rho$ to be zero inside the
volume $v_{\rm in}$ of the solute:
\be
\rho({\bf r})=n({\bf r})+ \omega({\bf r})\equiv 0 \label{eq:dens}
\ee
where ${\bf r}\in v_{\rm in}$. The direct interaction between the short-- and the
long--range density fluctuations in the solvent can be written in the simplest
possible way
\be
V[\omega,n]=c\int {\omega}({\bf r})\, n({\bf r})\,d{\bf r} \label{eq:direct}
\ee
where $c$ is the coupling constant. In what follows all the analytic derivations are
performed for generic $c$, however the numerical solutions of the corresponding
equations are presented for $c=0$ only. The case $c\neq 0$ we leave for the future
consideration.

The solute--solvent partition function is obtained as follows
\be Z=\int \mcl{D}\{\omega({\bf r})\}
\prod_{{\bf r}\in V_{\rm in}}\delta\left[\rho({\bf r})\right]e^{-\mcl{H}}
\label{eq:sum}
\ee
where
\bs
\beqn \mcl{H}=\mcl{H}_0[\omega,n]+V[\omega,n]+\int\, d\mbf{r} \,\varphi(\mbf{r})
\rho(\mbf{r}) \label{eq:hamilt} \\
\rho({\bf r})=n({\bf r})+\omega({\bf r})
\eeqn \es
Let us rewrite the constraints imposed by the Dirac $\delta$--function in
\eq{eq:sum} using the functional Fourier transform
\be
\prod_{{\bf r}\in v_{\rm rm}}\delta\left[\rho({\bf r})\right]=
\frac{1}{(2\pi)^{\mcl{N}}}\int \mcl{D}\{\psi({\bf r})\} \exp\left\{i \int_{v_{\rm
in}} \rho({\bf r})\, \psi({\bf r})\, d{\bf r} \right\} \label{eq:delta}
\ee
where $\mcl{N}$ is the number of points in the product $\prod\limits_{{\bf r}\in
v_{\rm in}}(...)$, and we have denoted $\mcl{D}\{\psi({\bf r})\}\equiv
\prod\limits_{{\bf r}\in v_{\rm in}}d\psi({\bf r})$. Now we can represent the
partition function $Z[\varphi({\bf r})]$ in (\ref{eq:sum}) as a functional integral
over an auxiliary field $\psi({\bf r})$:
\be \barr{r} \disp Z=\Xi^{-1} \int\mcl{D}\{\omega({\bf r})\}
\mcl{D} \{\psi({\bf r})\} e^{-S[\omega,n,\psi]}
\medskip \\ \disp S[\omega,n,\psi]=\mcl{H}[\omega,n]+
i\int_{v_{\rm in}} \rho({\bf r})\, \psi({\bf r})\, d{\bf r} \earr \label{eq:PF}
\ee
where $\Xi$ is some physically irrelevant constant which defines the normalization
of the partition function $Z$.

The partition function $Z$ with the action $S[\omega,n,\psi]$ defined in \eq{eq:PF}
is the key object for our computations. Such quantities as the total, $\rho({\bf
r})$, and the averaged, $n({\bf r})$, densities, the solvation free energy, $\Delta
G$, the free energy of interactions of different solute molecules in the solvent,
etc. can be calculated on the basis of the generating function $Z$.

\subsection{The mean density}

To define the mean (large scale) density $n(r)$, let us require
\be
\frac{\delta S[n({\bf r})]}{\delta n({\bf r})}=0 \label{eq:n}
\ee
In \eq{eq:n} all other fields except $n({\bf r})$ are supposed to be already
integrated over. This leads to the functional equation which can be written
explicitly as a set of two coupled integro--differential equations relating the
inner (${\bf r}\in v_{\rm in}$) and outer (${\bf r}\in v_{\rm out}$) regions of the
solute molecule:
\bs
\be
\barr{ll} \disp -a \Delta n({\bf r}) + \frac{\delta W(n({\bf r}))}{\delta n({\bf
r})} - 2 c n({\bf r}) - &  \medskip \\ \disp c \int_{v_{\rm in}} d{\bf
r}'\int_{v_{\rm out}} d{\bf r}'' \chi^{-1}_{\rm in}({\bf r},{\bf r}') \chi(|{\bf
r}'-{\bf r}''|) n({\bf r}'') + \int_{v_{\rm in}}d{\bf r}' \chi^{-1}_{\rm in}({\bf
r},{\bf r}') n({\bf r}') = 0  & \qquad \mbox{for ${\bf r}\in v_{\rm in}$} \earr
\label{eq:n_in}
\ee
and
\be
\barr{ll} \disp -a \Delta n({\bf r}) + \frac{\delta W(n({\bf r}))}{\delta n({\bf
r})} -c^2 \int_{v_{\rm out}} d{\bf r}'\chi(|{\bf r}-{\bf r}'|)n({\bf r}') +
\medskip \\ \disp c^2 \int_{v_{\rm in}} d{\bf r}'\int_{v_{\rm in}} d{\bf r}''
\int_{v_{\rm out}} d{\bf r}''' \chi(|{\bf r}-{\bf r}'|)\chi^{-1}_{\rm in}({\bf
r}',{\bf r}'') \chi(|{\bf r}''-{\bf r}'''|) n({\bf r}''') - & \medskip \\ \disp c
\int_{v_{\rm in}} d{\bf r}'\int_{v_{\rm in}} {\bf r}'' \chi(|{\bf r}-{\bf
r}'|)\chi^{-1}_{\rm in}({\bf r}',{\bf r}'') n({\bf r}'') = 0 & \qquad \mbox{for
${\bf r}\in v_{\rm out}$} \earr \label{eq:n_out}
\ee
\es

In the current paper we pay attention to the case $c=0$ only, i.e. when the {\it
direct coupling} between the short-- and long--ranged fluctuation is absent. In that
case the profiles of the mean density $n(r)$ for spherical solute molecules of
different sizes are given by the solutions of the simplified integro--differential
equation
\bs \beqn -a \Delta n({\bf r}) + \frac{\delta W(n({\bf r}))}{\delta n({\bf
r})}+ \int_{v_{\rm in}}d{\bf r}' \chi^{-1}_{\rm in}({\bf r},{\bf r}') n({\bf
r}') = 0 & \qquad \mbox{for ${\bf r}\in v_{\rm in}$} \label{eq:n_in_outa} \\
\disp -a \Delta n({\bf r}) + \frac{\delta W(n({\bf r}))}{\delta n({\bf r})}= 0 &
\qquad \mbox{for  ${\bf r}\in v_{\rm out}$} \label{eq:n_in_outb} \eeqn \es

We solve \eq{eq:n_in_outa}--\eq{eq:n_in_outb} numerically by using the simple
technical trick which allows to increase essentially the speed of the computations.
Namely, we have observed that the most CPU time is spent for the calculation of
$\chi^{-1}_{\rm in}({\bf r},{\bf r}')$ on the basis of \eq{chi_inv}. To get rid of
this part of computations we can multiply the \eq{eq:n_in_outa} by $\chi(|{\bf
r}-{\bf r}'|)$ and integrate over the whole space $v_{\rm tot}$ ($v_{\rm tot}=v_{\rm
in} \cup v_{\rm out}$), with the subsequent application of \eq{chi_inv}. Finally we
arrive at the following set of equations:
\be
\barr{rl} \disp n({\bf r}) + \int_{v_{\rm in}} d{\bf r}' \chi(|{\bf r}-{\bf r}'|)
\left(-a\Delta n({\bf r}') + \frac{\delta W(n({\bf r}'))}{\delta n({\bf r}')}\right)
=0 & \qquad \mbox{for ${\bf r}\in v_{\rm in}$} \medskip \\ \disp -a \Delta n({\bf
r}) + \frac{\delta W(n({\bf r}))}{\delta n({\bf r})}= 0 & \qquad \mbox{for  ${\bf
r}\in v_{\rm out}$} \earr \label{n_in_out}
\ee

\begin{figure}[ht]
\epsfig{file=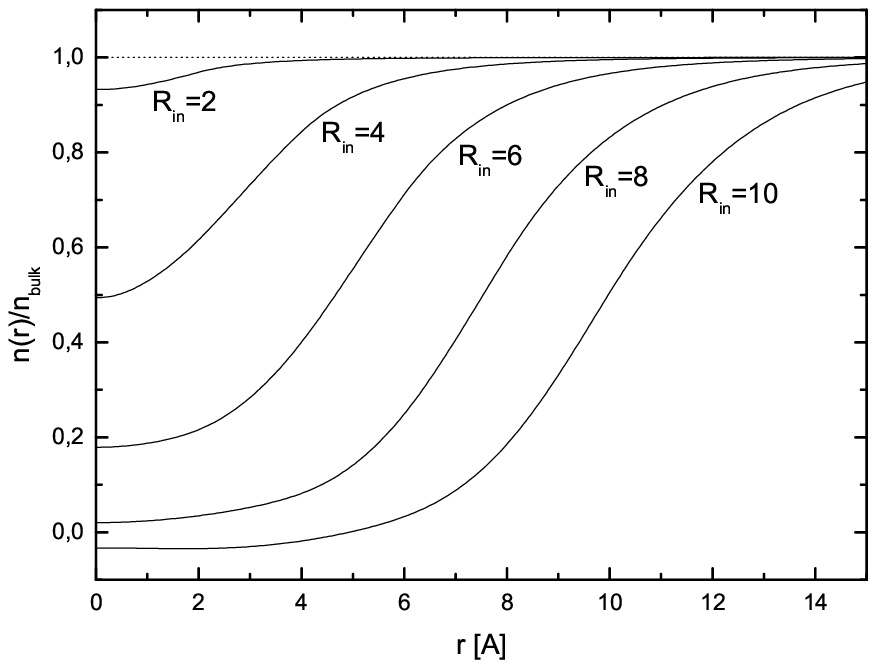,width=8cm} \epsfig{file=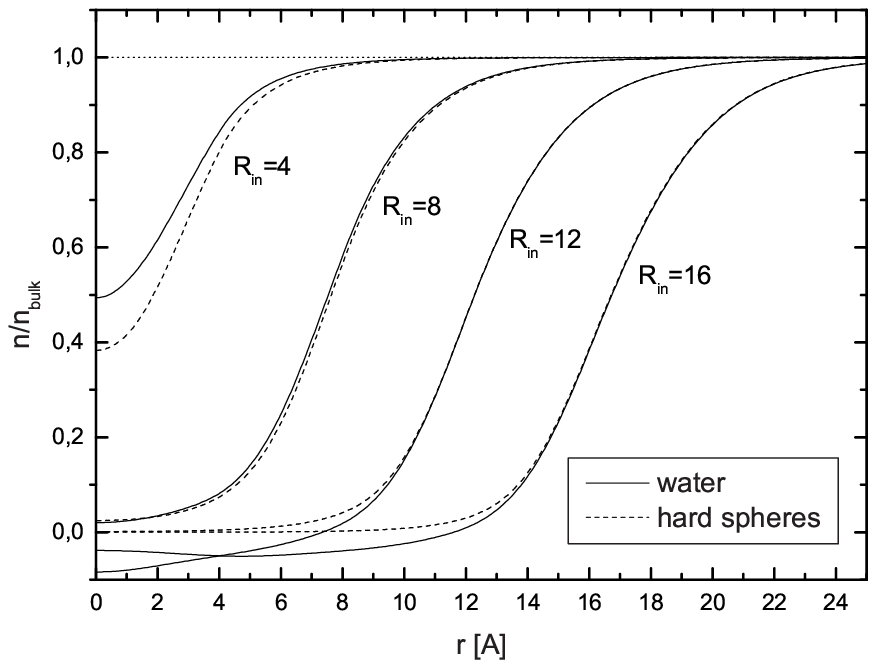,width=8cm}
\caption{Typical profiles of the mean density $n(r)$ around spherical solute
molecules of different radii $R_{\bf in}$. } \label{fig:2}
\end{figure}

The generic form of the mean density profile $n(r)$ for the spherical solute
molecule is shown in \fig{fig:2} for different (still arbitrary) values of the
constants $a$ and $b$ of Hamiltonian. The densities of the vapor ($n_1$) and bulk
($n_2$) phases are as follows: $n_1=0; \, n_2=0.7$. The numerical solutions of
\eq{n_in_out} are obtained with the correlation functions of hard spheres (dashed
line) and of water (solid line). The typical plots are shown in \fig{fig:2}.

The presence of the regions with the negative values of the mean densities in the
vapor phase should not confuse---remember that only the full density $\rho({\bf r})$
is a conserved value and should be zero inside the solute molecule, while the
average density $n({\bf r})$ is determined selfconsistently. The existence of the
regions with negative values of the long--ranged averaged field $n({\bf r})$
signifies that in these regions the fluctuations of the short--ranged field
$\rho({\bf r})$ are not symmetric, such that locally $\left<\omega\right>\neq 0$.
Such regions exist only inside the solute volume $v_{\rm in}$ and in the "physical"
volume $v_{\rm out}$ (i.e. outside the solute) the mean density is always positive.

The full density $\rho({\bf r})=n({\bf r})+\omega({\bf r})$ which minimizes the
functional \eq{eq:PF} satisfies the following set of equation
\be
\barr{llll}
\rho(\mbf{r}) & = & 0, & \qquad \mbox{for ${\bf r}\in v_{\rm in}$} \medskip \\
\rho(\mbf{r}) & = & \disp n(\mbf{r})-\int_{v_{\rm in}} d\mbf{r}'\int_{v_{\rm in}}
d\mbf{r}'' \chi(|\mbf{r}-\mbf{r}'|)\chi^{-1}_{\text{in}}(\mbf{r}',\mbf{r}'')
n(\mbf{r}'')+ & \medskip \\ & & \disp c\int_{v_{\rm in}} d\mbf{r}'\int_{v_{\rm in}}
d\mbf{r}'' \int_{v_{\rm in}} d\mbf{r}''' \chi(|\mbf{r}-\mbf{r}'|)
\chi^{-1}_{\text{in}}(\mbf{r}',\mbf{r}'') \chi(|\mbf{r}-\mbf{r}'''|) n(\mbf{r}''')-
& \medskip \\ & & \disp c\int_{v_{\rm out}} d\mbf{r}'\chi(|\mbf{r}-\mbf{r}'|)
n(\mbf{r}') & \qquad \mbox{for ${\bf r}\in v_{\rm out}$} \earr \label{eq:full_dens}
\ee

The equations \eq{eq:full_dens} with $c=0$ again can be rewritten in the form which
excludes the function $\chi^{-1}_{\text{in}}(\mbf{r},\mbf{r}')$ from the
corresponding expressions.

\subsection{The solvation free energy}

The free energy of solvation $\Delta G$ is defined as the energy necessary to
transport the solute molecule from its environment in the solvent to the vacuum. The
partition functions of solvent samples of sufficiently large volume $V_{\rm
solvent}$ containing the solute molecule inside can be straightforwardly written on
the basis of \eq{eq:sum}:
\be
Z_{\rm solvent-solute} = \int_{v_{\rm solvent}} \mathcal{D}\{\omega({\bf
r})\}\prod_{{\bf r}\in v_{\rm solute}}\delta\left[\rho({\bf
r})\right]e^{-\mathcal{H}} \label{eq:st_sum}
\ee
where $\mathcal{H}$ is given by \eq{eq:hamilt}. Conventionally the solvation free
energy $\Delta G$ is written as follows
\be \Delta G = -\ln \frac{Z_{\rm solvent-solute}}{Z_{\rm solvent\; only}}
\label{eq:deltaG}
\ee
where $Z_{\rm solvent\; only}$ is the partition function of the pure solvent
(without the solute molecules). At first glance it seems naturally to write $Z_{\rm
solvent\; only}$ simply as
\be Z_{\rm solvent\; only} = \int_{v_{\rm solvent}} \mathcal{D}
\{\omega({\bf r})\}\,e^{-\mathcal{H}} \label{eq:false}
\ee
However the expression \eq{eq:false} being used in \eq{eq:deltaG} leads to the
divergence of the solvation free energy $\Delta G$ in practical computations based
on the approach developed in \eq{eq:sum}--\eq{n_in_out}. The formal reason for such
a divergence deals with the occurrence of uncompensated infinite product of Gaussian
integrals \eq{eq:PF} in the ratio $Z_{\rm solvent-solute}/Z_{\rm solvent\; only}$ in
\eq{eq:deltaG}. The physical origin of such a divergence is due to the mixture of
different statistical ensembles associated with the partition functions
\eq{eq:st_sum} and \eq{eq:false}. Namely, when writing $Z_{\rm solvent-solute}$ as
in \eq{eq:st_sum} and imposing the $\delta$--function constraint, we fix some
particular value of the field $\rho$ within the solute volume $v_{\text{in}}$ what
means that \eq{eq:st_sum} is the partition function of the canonical ensemble (with
respect to the density $\rho$). At the same time, for $Z_{\rm solvent\; only}$
written in the form \eq{eq:false}, we allow any density of the field $\rho$ inside
the solute volume. Hence, \eq{eq:false} is the partition function of the grand
canonical ensemble.

The regularization of eq.\eq{eq:false} is based on the probabilistic consideration
of the solvation free energy \cite{chand2}. Rewrite $\Delta G$ in \eq{eq:deltaG} as
follows
\be
\Delta G=-\ln \frac{Z_v(0)}{\sum_{N\geq 0} Z_v(N)}
\ee
where $Z_v(N)$ is a partition function of $N$ molecules inside the volume $v_{\rm
in}$. The continuous analog of the last expression appears when passing from $N$ to
the average number of molecules $\tilde{n}=N/v_{\rm in}$. Simultaneously we require
the net density $\rho$ to be equal $\tilde{n}$ inside $v_{\rm in}$ and replace the
summation over $N$ by the integration over $\tilde{n}$. Now we can rewrite the
normalization partition function $Z_{\rm solvent\; only}$ in the following
"regularized" form
\be
Z_{\rm solvent\; only} = \int_0^{\infty} d\tilde{n} \int_{v_{\rm solvent}}
\mathcal{D}\{\omega({\bf r})\}\prod_{{\bf r}\in v_{\rm solute}}\delta\left[\rho({\bf
r})-\tilde{n}\right]\,e^{-\mathcal{H}} \label{eq:true}
\ee
Instead of this form we use an approximation
\be
Z_{\rm solvent\; only} = \int_{v_{\rm solvent}} \mathcal{D}\{\omega({\bf
r})\}\prod_{{\bf r}\in v_{\rm solute}}\delta\left[\rho({\bf
r})-\bar{n}\right]\,e^{-\mathcal{H}} \label{eq:true_spi}
\ee
obtained from \eq{eq:true} by the stationary phase integral calculation technic.
Thus, the definition \eq{eq:deltaG} with the partition functions given by
\eq{eq:st_sum} and \eq{eq:true_spi} is very natural, justified physically and does
not contain any divergences. It is important to notice that the computation of the
equilibrium density based on $Z_{\rm solvent\; only}$ given by \eq{eq:true} is the
bulk density, for which the value of corresponding effective Hamiltonian is strictly
zero.

The solvation free energy $\Delta G$ for any solute molecule is computed on the
basis of the equilibrium density profile $n_{\rm eq}({\bf r})\equiv n({\bf r})$
calculated from Eqs.\eq{eq:n_in}--\eq{eq:n_out}. Substituting $n_{\rm eq}({\bf r})$
back into the Hamiltonian \eq{eq:hamilt} and taking into account \eq{chi_inv}, we
can rewrite the solvation free energy $G\equiv \mcl{H}\{n_{\rm eq}\}$ as a sum of
three terms:
\be
\Delta G_{\rm solv}(v_{\rm in})= \Delta G_{\rm corr} + \Delta G_{\rm surf} + \Delta
G_{\rm int} \label{eq:G_H}
\ee
where
\be
\barr{lll} \Delta G_{\rm corr} & = & \disp \frac{1}{2}\int_{v_{\rm in}} n_{\rm
eq}({\bf r}) \chi_{\rm in}^{-1}({\bf r},{\bf r}') n_{\rm eq}({\bf r}') \, d{\bf r}
d{\bf r}' \medskip \\ \Delta G_{\rm surf} & = & \disp \frac{a}{2} \int (\nabla
n_{\rm eq}({\bf r}))^2 d{\bf r} \medskip \\ \Delta G_{\rm int} & = & \disp \int
W\left(n_{\rm eq}({\bf r})\right) d{\bf r} \label{eq:G_parts} \earr
\ee

\begin{figure}[ht]
\epsfig{file=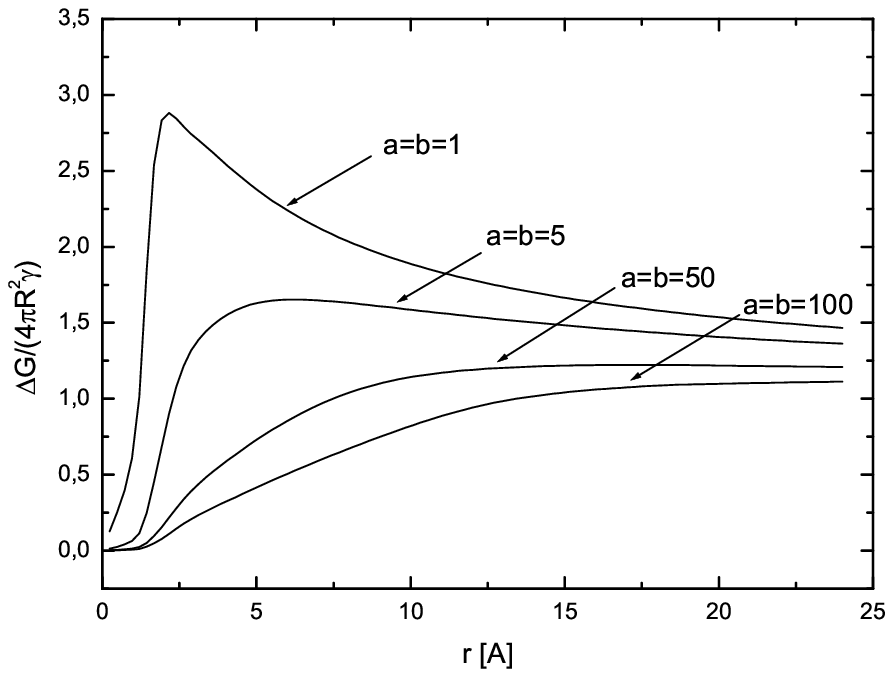,width=8cm} \epsfig{file=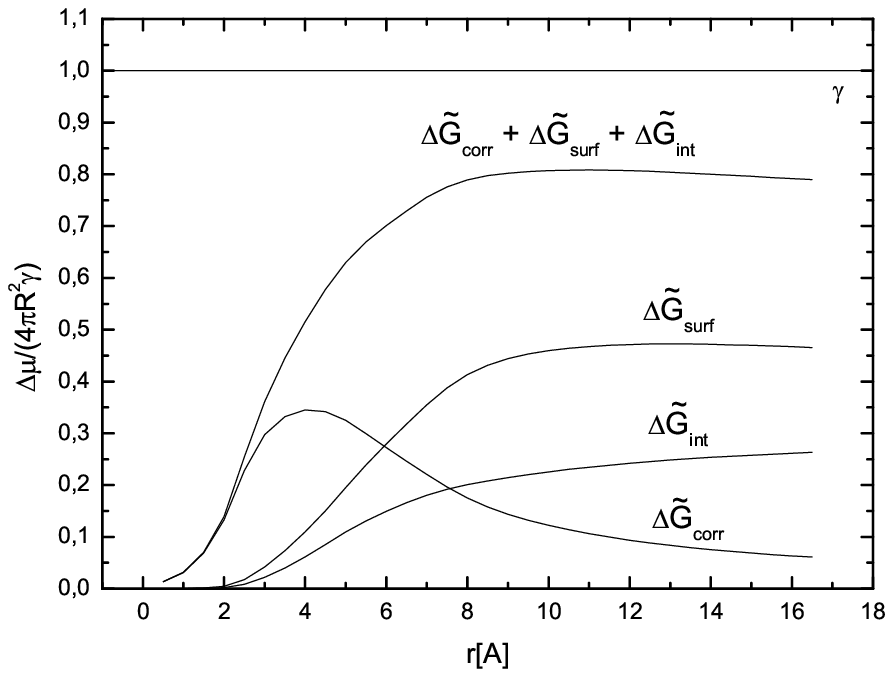,width=8cm}
\caption{Typical behavior of the solvation free energy for different choice of
parameters $a$ and $b$ in the Hamiltonian (left); Contributions to the normalized
solvation free energy $\Delta \tilde{G}_{\rm solv}(R_{\rm in})$ of a spherical
solute from $\Delta \tilde{G}_{\rm corr}$, $\Delta \tilde{G}_{\rm surf}$ and $\Delta
\tilde{G}_{\rm int}$ for $a=9,\, b=30$ (right).} \label{fig:3}
\end{figure}

As in the computations of the mean density $n(r)$, here we
consider the case $c=0$ only. In that case \eq{eq:G_H} can be
rewritten in the following form \be \Delta G_{\rm solv}(v_{\rm
in})=\int_{v_{\rm in}} \left(W(n)-\frac{n}{2}\frac{\delta
W}{\delta n}\right) d\mbf{r} \label{eq:G} \ee The contributions to
the normalized solvation free energy \be \Delta \tilde{G}_{\rm
solv}=\frac{\Delta G_{\rm solv}}{4 \pi r_{\rm in}^2}
\label{eq:tildeG} \ee of a spherical solute molecule of radius $R$
are shown in Fig.\ref{fig:3} for each term $\{\Delta
\tilde{G}_{\rm corr}$, $\Delta \tilde{G}_{\rm surf}$, $\Delta
\tilde{G}_{\rm int}\}= \{\Delta G_{\rm corr}$, $\Delta G_{\rm
surf}$, $\Delta G_{\rm int}\}/(4\pi R_{\rm in}^2)$ separately, as
well as for the sum $\Delta \tilde{G}_{\rm solv}=\Delta
\tilde{G}_{\rm corr}+\Delta \tilde{G}_{\rm surf}+\Delta
\tilde{G}_{\rm int}$. The diameter of the solvent molecule in
\fig{fig:3} is set to 1.

The solvation free energy of a spherical solute molecule in absence of electrostatic
interactions \cite{chand1,chand2,chand3} is proportional (as expected) to the volume
of solute molecule for sufficiently small sizes $R_{\rm in}$ (of order of the
correlation length in the solvent) and tends to be proportional to the surface area
for large $R_{\rm in}$. We clearly see the non-monotonic behavior of the solvation
energy upon the size of a solute molecule \footnote{Let us stress that such behavior
we have found in the absence of the direct interactions between small-- and
large--length scales (i.e. for $c=0$ in \eq{eq:direct}).} reported in some works
(see, for example, \cite{chand4}).

\subsection{Adjustment of the parameters of the Hamiltonian}

The free parameters $a$ and $b$ of the Hamiltonian \eq{eq:PSH}--\eq{eq:GLP} are
chosen from two basic requirements:

i) The theoretically computed normalized solvation free energy $\Delta
\tilde{G}_{\rm solv} =\Delta G_{\rm solv}/(4\pi R^2)$ of a spherical solute molecule
of a radius $R$ (see \eq{eq:tildeG}) reproduces the corresponding dependence $\Delta
\tilde{G}_{\rm solv}(R)$ obtained in the Monte--Carlo simulations of \cite{chand4};

ii) The theoretically computed full density $\rho(r|R)$ for few sizes $R$ of
spherical solute molecules \eq{eq:full_dens} reproduces the behavior $g(r)$ of the
correlation function in vicinity of the spherical solute of radius $R$ extracted
from the Monte--Carlo simulations \cite{chand4}.

\begin{figure}[ht]
\epsfig{file=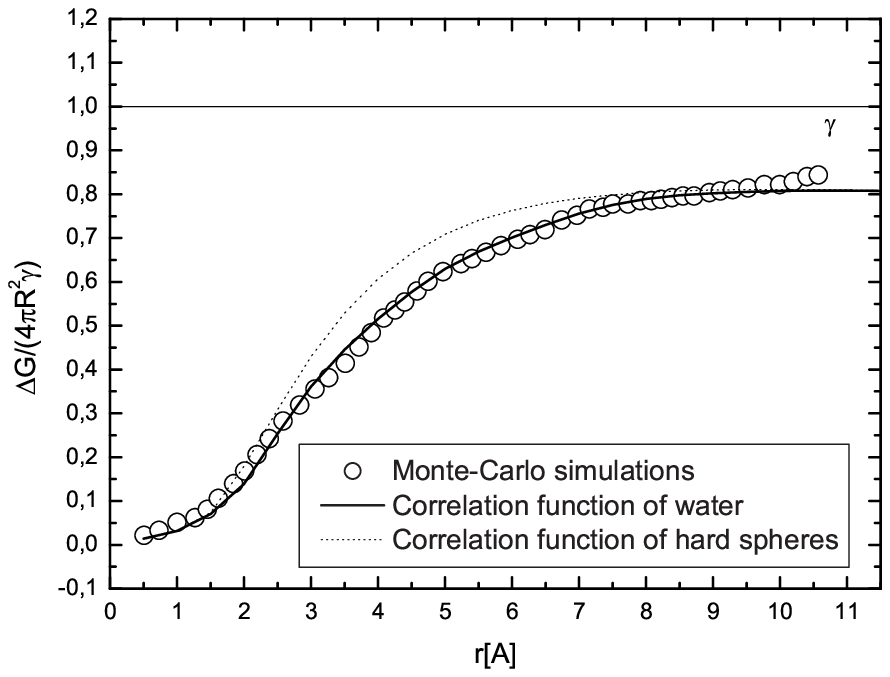,width=8cm} \epsfig{file=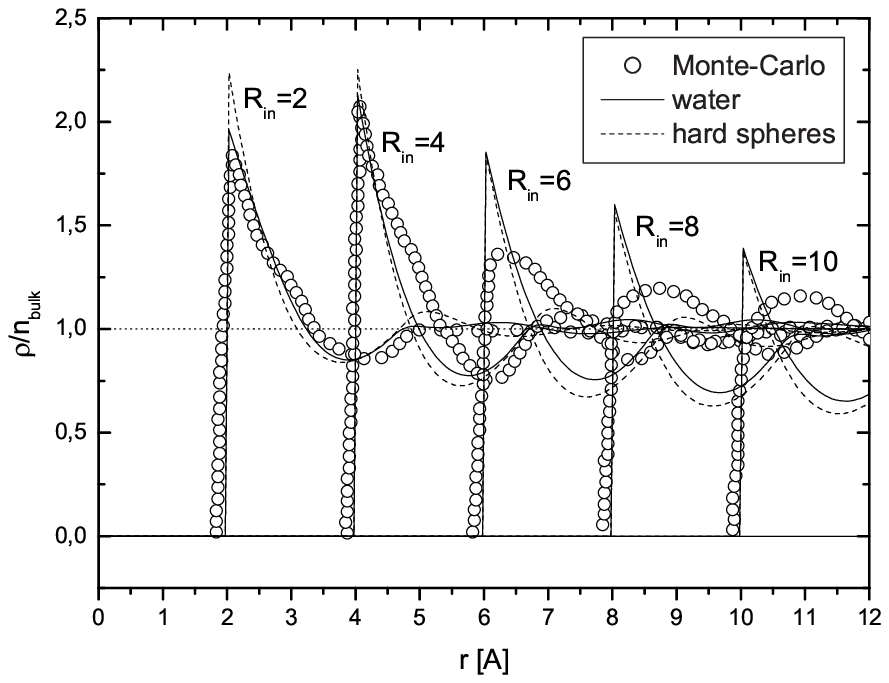,width=8cm}
\caption{Solvation free energy \eq{eq:G} of a spherical solute {\it vs} Monte--Carlo
simulations \cite{chand4} (left); Profiles of the full density $\rho(r)$ around
spherical solutes of different radii $R$ {\rm vs} Monte--Carlo simulations
\cite{chand4} for the correlation function $g(r)$.} \label{fig:G_rho}
\end{figure}

The corresponding results are shown in \fig{fig:G_rho} for the following choice of
the parameters: $a=9;\, b=30$.

The dashed and solid lines represent the results obtained with the correlation
functions of hard spheres and of the water correspondingly. The solvation energy
$\Delta \tilde{G}_{\rm solv}$ is represented in \fig{fig:G_rho} in dimensionless
units $\Delta \tilde{G}_{\rm solv} /\gamma$, where $\gamma$ is the surface tension
of the flat vapor--water interface. The numerical value $\gamma\approx 72$ mJ/m$^2$
is taken from \cite{chand4}.

The solvation energies $\Delta G_{\rm solv}$ and $\Delta \tilde{G}_{\rm solv}$
computed directly using \eq{eq:G} and \eq{eq:tildeG} can be converted to the proper
units by means of the following scaling coefficients:
\be
\left\{\barr{l} \disp \Delta G_{\rm solv},\, [\mbox{dimensionless units}] \times
0.2479 = \Delta G_{\rm solv},\, [\mbox{kcal/mol}] \medskip \\
\disp \Delta \tilde{G}_{\rm solv},\, [\mbox{dimensionless units}] \times
\frac{0.2479}{(2.78)^2} = \Delta \tilde{G}_{\rm solv},\, [\mbox{kcal/(mol \AA$^2$)}]
\earr \right. \label{eq:coeff}
\ee
These transfer coefficients remain unchanged in all further computations. In
\fig{fig:3} and \fig{fig:G_rho} the radius of a solute molecule is measured in
Angstroms.

\section{Results and discussions: Solvation of objects of various shapes}
\label{disc}

The current mean--field--type theory of solvation is applied to computation of
solvation of neutral molecules (alkanes), as well as to the computation of the free
energy of interactions of separated objects (spheres).

\subsection{Solvation of alkanes}

The parameters of the theory and the scaling coefficient are adjusted to satisfy the
Monte--Carlo simulations of solvation of hard spheres and the corresponding profile
of the solvent density near the solvated object (see the previous Section) and are
not tuned anymore. In particular, we use the same parameters to compute analytically
the solvation free energy of alkanes. As it is shown below, we find very good
coincidence of our computed values with the predictions of the "Scaled Particle
Theory" which uses the parameters tuned especially to alkanes \cite{irisa}.

The mentioned coincidence needs some elucidations. There is a viewpoint
\cite{lee,widom,levy} supported by numerical computations that one can split the
interaction between the solute and the solvent into two parts: i) the free energy,
$\Delta G_{\rm cav}$, of a "cavity formation", and ii) the dispersion (attractive)
part of the Van-der-Waals interactions, $\Delta G_{\rm disp}$. The numerical methods
involving the "thermodynamic integrations" allow to compute both the contributions,
$\Delta G_{\rm cav}$ and $\Delta G_{\rm disp}$ separately \cite{levy}. These
contributions can be also computed separately in the frameworks of the "Scaled
Particle Theory" (SPT) developed in \cite{spt,spt1}. The corresponding description
of alkanes has been undertaken in \cite{irisa} where the authors have reported very
good quantitative agreement of the sum $\Delta G_{\rm cav}+\Delta G_{\rm disp}$ with
the experimentally measured solvation free energy.

The model described in the previous sections of our work considers the solute as an
object bounded by hard walls and hence takes into account only the "cavitation" part
of the solvation free energy. The attractive part of the Van-der-Waals interactions
is not yet taken into account. However we do not see any principal obstacles in
adding the dispersion part of solute--solvent interactions to our model. Moreover,
we can "smear" the $\delta$--functional constraint in \eq{eq:delta}, releasing more
realistic form of the Van-der-Waals potential. The corresponding computations are in
progress and will be reported in a forthcoming publication \cite{sineta}.

The comparison of our predictions for the cavitation part of the solvation free
energy of alkanes with the corresponding contribution $\Delta G_{\rm cav}$ extracted
from the papers \cite{irisa} and \cite{pais} is shown in the table below.

\begin{table}[ht]
\begin{tabular}{|l||c|c|c|} \hline
alkanes & $\Delta G_{\rm cav}$, kcal/mol (our model) & $\Delta G_{\rm cav}$,
kcal/mol (from \cite{irisa}) & $\Delta G_{\rm cav}$, kcal/mol (from \cite{pais}) \\
\hline \hline ${\rm CH_4}$ & 5.69 & 5.61 & 5.36 \\
\hline ${\rm C_2H_6}$ & 7.54 & 7.53 & 7.15 \\ \hline ${\rm C_3H_8}$ & 9.07 & 9.13 & \\
\hline ${\rm C_4H_{10}}$ & 10.52 & 10.8 & \\ \hline ${\rm C_5H_{12}}$ & 11.96 & 12.8 & \\
\hline ${\rm C_6H_{14}}$ & 13.24 & 14.8 & 14.22 \\ \hline
\end{tabular}
\caption{Comparison of the "cavity formation" contribution, $\Delta G_{\rm cav}$, to
the solvation free energy of alkanes computed on the basis of our model with the
corresponding data extracted from the papers \cite{irisa} and \cite{pais}.}
\label{tab:1}
\end{table}

Let us remind once more that in our model the parameters of the Hamiltonian $a=9$,
$b=30$ and the scaling factors \eq{eq:coeff} are tuned on the basis of the
Monte--Carlo simulations of solvation of hard spheres {\it without any additional
adjustment} to alkanes. The volume $v_{\rm in}$ occupied by the alkane molecules we
compute using the following prescription. We take the coordinates of centers of all
atoms from a standard database, then we draw the spheres with the effective radii
3.5 \AA~ for all carbon atoms (the length of the C--O bond in water) and 3.05 \AA~
for all hydrogen bonds (the length of the O--H bond in water) and find $v_{\rm in}$
as the union of the corresponding spherical volumes. The corresponding numerical
values of C--O and O--H bonds are extracted from the Monte--Carlo (MC) simulations
of the sytemem consisting of 216 water molecules and one methane molecule and
subsequent computations of the pairwise correlation functions [C(methane)--O(water)]
and [H(methane)--O(water)], \cite{ozrin}. The same parameters can be extracted using
the force field developed in \cite{halgren}.

As it has already been mentioned above, the dispersion part of the Van-der-Waals
(VdW) interaction can be easily taken into account. In the presence of the
attractive contribution to the VdW interactions, we should add an extra term $U_{\rm
disp}$ to the mean--field potential \eq{eq:PSH}, where
$$
U_{\rm disp} = \int U({\bf r})\, n({\bf r}) d{\bf r}
$$
and $U({\bf r})$ is the attractive part of the VdW interactions.

In the presence of the potential $U({\bf r})$ the equations
\eq{eq:n_in_outa}--\eq{eq:n_in_outb} for the equilibrium mean density are modified
in the following simple way: \eq{eq:n_in_outa} reminds without changes, while the
equation \eq{eq:n_in_outb} acquires an additional contribution from the attractive
part of the potential $U({\bf r})$ in the region ${\bf r}\in v_{\rm out}$:
\be
-a \Delta n({\bf r}) + \frac{\delta W(n({\bf r}))}{\delta n({\bf r})} + U({\bf r}) =
0 \label{eq:u(r)}
\ee

\subsection{Solvation of cylinders and the free energy of interactions of two
separated spheres}

\begin{figure}[ht]
\epsfig{file=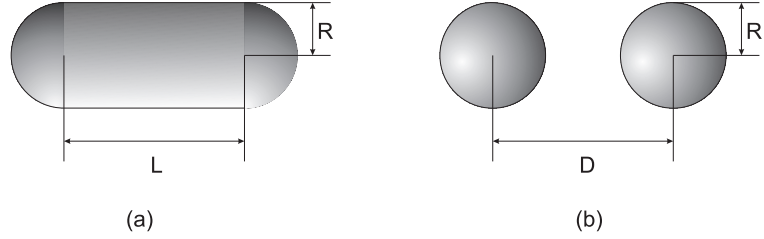,width=10cm} \caption{Schematic pictures: of a solvated
cylinder ended by two semi-spheres (a), and of two spherical solute molecules (b).}
\label{fig:5}
\end{figure}

In this section we consider the cavitation part of the solvation free energy of
molecules of cylindrical geometry, as well as of the free energy of interaction of
two hard spheres in the fluctuating media (i.e. in water). Schematically the studied
systems are shown in \fig{fig:5}. The purpose of this section is two--fold.

1. First of all, we compute the depth of the "skin layer" of the solvated object.
The naive computation of the solvation free energy (as well as of the free energy of
interactions) of large molecules like ligands and proteins demands essential
computational resources. Actually, according to \eq{n_in_out} one has to take into
account the interactions of the fluctuating density field $n({\bf r})$ with all
points inside the solvated object $v_{\rm in}$. However it is clear that due to the
hard wall constraint the fluctuating field $n({\bf r})$ cannot penetrate deep inside
the body of a solute. Considering the solvation of cylindric molecules of different
widths and lengths, we study the penetration depth of the density field $n({\bf r})$
inside the solute molecule.

\begin{figure}[ht]
\epsfig{file=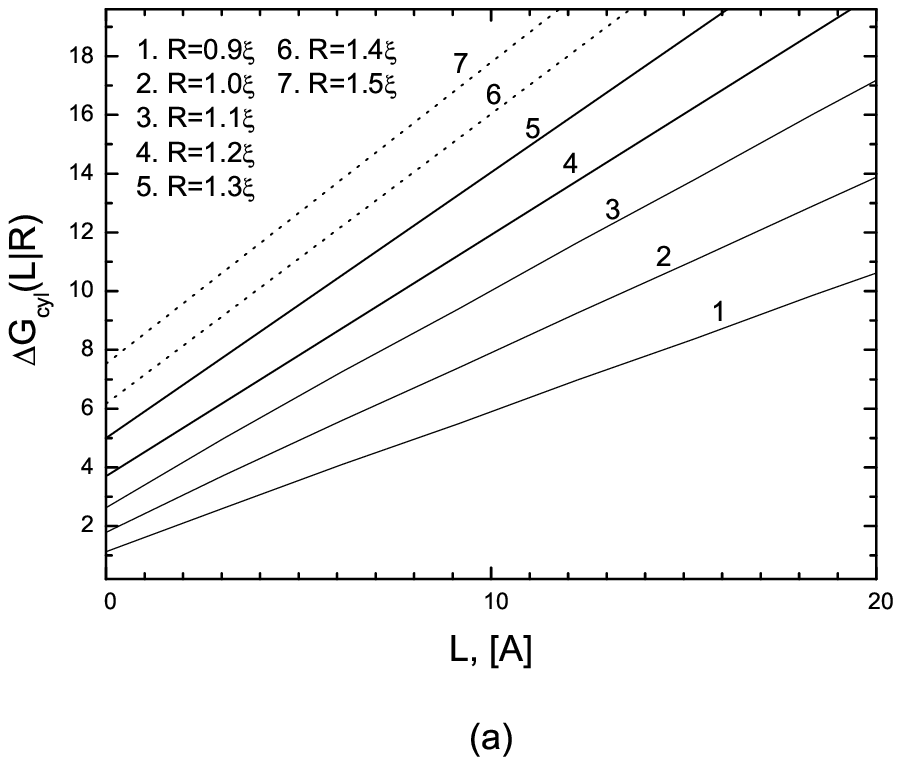,width=8.1cm} \epsfig{file=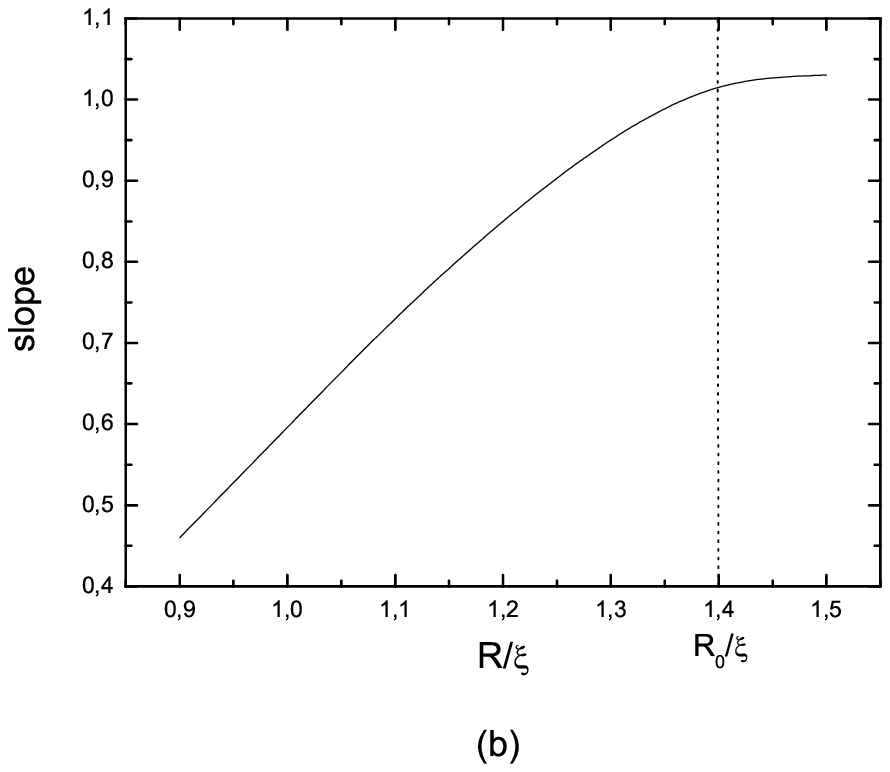,width=8cm} \caption{(a)
Solvation free energy $\Delta G_{\rm cyl}(L|R)$ of a cylinder as a function of its
length $L$ for different $R$ in the range $R=0.9...1.5$; (b) Slope of $\Delta G_{\rm
cyl}(R)$ as a function of $R$ at $L=12$ \AA.} \label{fig:6}
\end{figure}

The dependence of the solvation free energy $\Delta G_{\rm cyl}(L|R)$ upon its
length $L$ for two different radii $R$ is shown in \fig{fig:6}a.

\begin{figure}[ht]
\epsfig{file=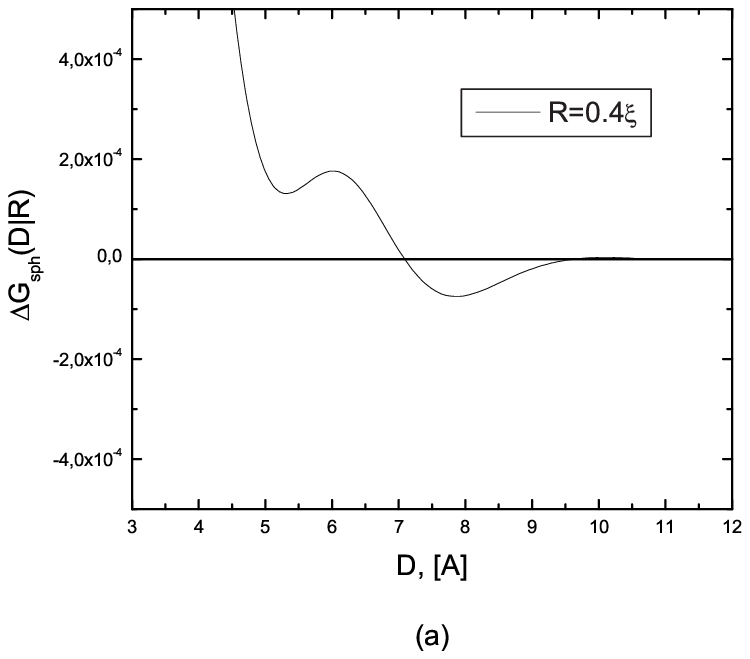,width=5.2cm} \epsfig{file=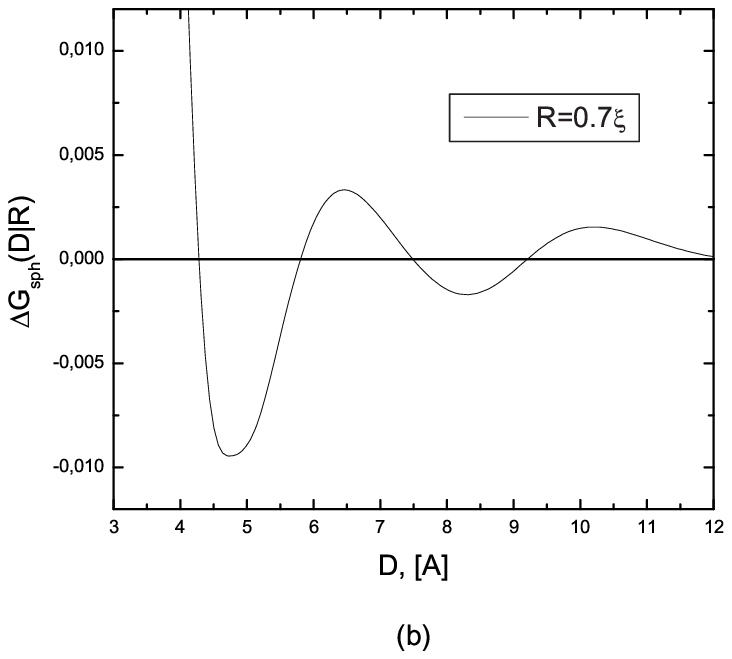,width=5cm}
\epsfig{file=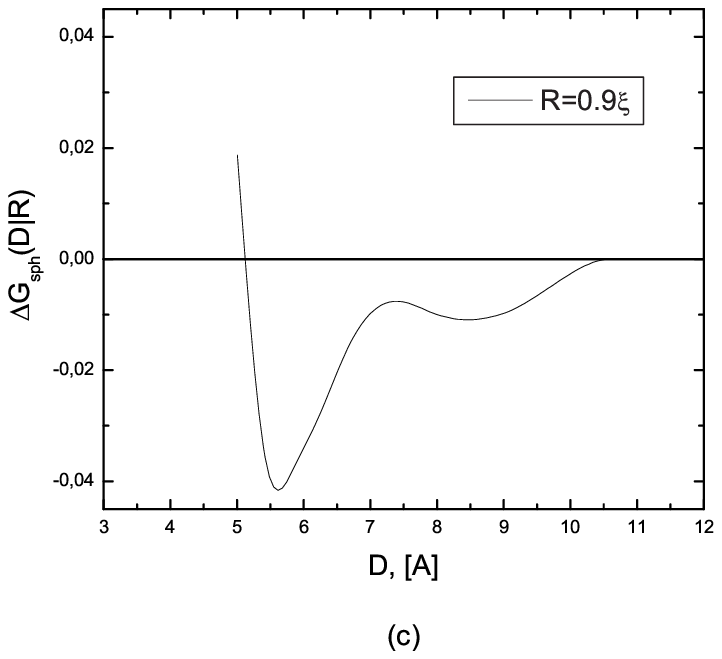,width=5cm} \caption{Free energy of interactions $\Delta
G_{\rm sph}$ of two spherical solute molecules as a function of the distance between
their centers $D$ for $R=0.4\xi$ (a); $R=0.7\xi$ (b) and $R=0.9\xi$ (c).}
\label{fig:7}
\end{figure}

As one sees, from \fig{fig:7}b, the slope of the curves in \fig{fig:6} becomes
constant approximately for $R>R_0\approx 1.4\xi=3.9$ \AA. It means that for $R>R_0$
the fluctuating density field penetrates inside the body of the solute to some fixed
length only (to the depth of the "skin layer"). Hence, we may not account of all
internal points of the solute located into its body beyond the skin layer
$R_0\approx 3.9$ \AA.

2. Secondary, we investigate the free energy of interaction $\Delta G_{\rm
sph}=\Delta G_{\rm sph\,int} (D|R)-\Delta G_{\rm sph\,solv}(D\to\infty|R)$ of two
spherical solute molecules, each of radius $R$, upon the distance $D$ between their
centers. The free energy of interaction, or "the mean force potential"
\cite{dill,shi} $\Delta G_{\rm sph}(D|R)$, with the subtracted solvation energies of
two solitary spheres, is computed for the hard--core potential, $U(D)$, where
$$
U(D)=\left\{\barr{ll} \infty & \mbox{for $D\leq 2R$} \medskip \\ 0 & \mbox{for $D>
2R$} \earr\right.
$$
i.e. we again take into account only the cavitation part of the free energy of
interactions, neglecting the dispersion (attractive) part of the direct
Van-der-Waals interactions between the impenetrable spheres of radius $R$.

Even for such simplified situation we have found that the behavior of $\Delta G_{\rm
sph}(D|R)$ on $D$ is qualitatively different for different values of $R$ compared to
$\xi$ \footnote{Let us remind that $\xi=2.78$ \AA~ is the location of the first
maximum of the correlation function of water.}. Let us stress once more that $\Delta
G_{\rm sph}$ is the free energy of interaction between two spheres with the
subtracted solvation energies of two solitary spheres $\Delta G_{\rm
sph\,solv}(D\to\infty|R)$. The observed dependencies are shown in \fig{fig:7}.

For all radii, $R=0.4\xi$ \AA~, $0.7\xi$ \AA~ and $0.9\xi$ \AA~ we reproduce the
non-monotonic dependence $\Delta G_{\rm sph}(D)$ on $D$. This non-monotonicity is the
manifestation of the oscillatory behavior of the water correlation function. Such
behavior has been found earlier in direct numerical simulations consistent with the
predictions of the "Scaled Particle Theory" (see \cite{dill} and the references
therein), and on the basis of estimation of the solvent accessible surface area
\cite{shi}. The account of the Van-der-Waals attractive tail in the potential $U(D)$
can essentially change the behavior $\Delta G_{\rm sph}(D)$, however the
non-monotonic behavior $\Delta G_{\rm sph}(D)$ will still hold.

Comparing \fig{fig:7}a and \fig{fig:7}c, we see that the dependence $\Delta G_{\rm
sph}(D)$ is sensitive (even qualitatively) to the size $R$ of interacting spheres.
The effective attraction between smaller spheres is weaker than that of larger
spheres. This effect can be easily understood. For sufficiently large spheres,
$R\gtrsim\xi$, there is an penalty in the free energy of keeping two solute
molecules close to each other, which is the "classical manifestation" of the
Casimir--type effect. At the same time, the sufficiently small solute molecules
(with $R \lesssim \xi$) are located in spontaneously created fluctuational cavities
of size $\xi$ in water and the penalty for expelling the water from the volume
occupied by two solute molecules close to each other is essentially smaller than
that for large solutes. Moreover, one can see from \fig{fig:7}a that the mean force
potential has the attractive (negative) part corresponding to the second water shell
only. So, the two sufficiently small spherical solutes can form a bound state
separated by one water shell. The found behavior qualitatively coincides with the
results of the work \cite{dill}, where the authors have studied the "potential of
the mean force" acting between two spherical solute molecules.

\subsection{Electrostatics in fluctuating dipolar environment}

There are different ways to implement the electrostatic interactions into our
self--consistent mean--field description. In many works it is assumed that the
dielectric permittivity of the solvent, $\varepsilon$, linearly depends on the
solvent density, $n({\bf r})$, i.e., $\varepsilon({\bf r}) = 1 + \alpha\, n({\bf
r})$. In our approach we get rid of such supposition and follow the self--consistent
scheme where the solvent is considered as a "gas" of dipoles. Our method
ideologically is very close to the consideration of the Debye screening in
electrically neutral plasma \cite{ll_5}. However, to the contrast with plasma, in
our case the positive and negative ions are connected by holonomic constraints in
pairs, forming the short--ranged dipoles. The screening in the system of extended
objects was a subject of many investigations. The description, most appropriate for
our goals, has been developed in the appendix of the paper \cite{khkh}.

Below we derive the basic set of equations describing simultaneously electrostatic
and hydrophobic interactions in the fluctuating media. The detailed analysis of the
obtained equations together with the numerical computation of the free energies of
solvation and interactions of charged molecules, will be the subject of the
forthcoming publication \cite{sineta}, while here we analyze briefly the limiting
case, in which the proposed theory coincides with the electrostatics in the
continuous media with some effective density--dependent permittivity
$\varepsilon({\bf r})$.

Let us introduce the quantities:
\begin{description}
\item $\rho_{\rm  mol}^q({\rm r},\tau)$ -- the charge distribution of the solvent
molecule with the orientation $\tau$ and with the center of mass situated at the
point ${\bf r}$. Outside of the molecule $\rho_{\rm mol}^q({\bf r},\tau)=0$;
\item $\rho_{\rm solute}^q({\bf r})$ -- the charge distribution of the solute;
\item $\phi_{\rm eff}({\bf r})$ -- the effective mean electrostatic field.
\end{description}

Sticking to the mean--field description developed in \cite{khkh}, we can place each
solvent molecule in a self--consistent field  electrostatic field, $\phi(\mbf{r})$,
averaged over the volume of this solvent molecule. So, we have:
\be
\phi(\mbf{r},\tau)=\int \phi_{\rm eff}({\bf r}')\rho_{\rm mol}^q({\bf r}'-{\bf
r},\tau)\, d{\bf r}' \label{eq:el1}
\ee

The electrostatic potential $\phi_{\text{eff}}(\mbf{r})$ is determined by the
Poisson equation
\be
\Delta \phi_{\text{eff}}(\mbf{r})=-\frac{4\pi}{\varepsilon_{\text{solute}}}
\rho_{\text{solute}}^{q}(\mbf{r})-\mu(\mbf{r}) \label{eq:el2}
\ee
where $\mu(\mbf{r})$ is the media charge density
\be
\mu(\mbf{r})=\int\int \left(e^{-\beta
\phi(\mbf{r}',\tau)}-1\right)\rho_{\text{mol}}^{q}(\mbf{r}-\mbf{r}',\tau)\rho(\mbf{r}')
\frac{d\Omega_\tau}{4\pi} d^{3} \mbf{r}' \label{eq:el3}
\ee
and $\rho(\mbf{r})=n(\mbf{r})+\omega(\mbf{r})$ is the solute density. Neglecting the
short--range fluctuations, $\omega(\mbf{r})$ we may write in the mean--field
approximation
\be
\mu(\mbf{r})=\int\int \left(e^{-\beta
\phi(\mbf{r}',\tau)}-1\right)\rho_{\text{mol}}^{q}(\mbf{r}-\mbf{r}',\tau)
n(\mbf{r}') \frac{d\Omega_\tau}{4\pi} d^{3} \mbf{r}' \label{eq:el4}
\ee

The full Hamiltonian of the system reads now:
\be
\barr{lll}
H[n(\mbf{r}),\phi_{\text{eff}}(\mbf{r})] & = & \disp
\int d\mbf{r}\,\left(\frac{a}{2}(\nabla n(\mbf{r}))^2 +
W(n)\right) +\frac{1}{2}\int_{v_{\text{in}}}
d\mbf{r}\,d\mbf{r}'\,n(\mbf{r})\chi^{-1}_{\text{in}}(\mbf{r},\mbf{r}')
n(\mbf{r}') \medskip \\ & + &
\disp \lambda \int d\mbf{r} \,(\nabla \phi_{\text{eff}}(\mbf{r}))^2
 +\int d\mbf{r}\, \varphi(\mbf{r})\left(\Delta
\phi_{\text{eff}}(\mbf{r})+\frac{4\pi}{\varepsilon_{\text{solute}}}
\rho_{\text{solute}}^{q}(\mbf{r})+\mu(\mbf{r})\right) \label{eq:el_ham}
\earr
\ee
Let us stress that according to \cite{ll_8} we should minimize the functional action
of hydrophobic interactions together with the one of electrostatic field, while the
Poisson equation \eq{eq:el2}) is considered as a constraint and hence enters in the
action with the functional Lagrange multiplier $\varphi({\bf r})$.

Minimizing the functional $H[n(\mbf{r}),\phi_{\text{eff}}(\mbf{r})]$ with respect to
the fields $\varphi(\mbf{r})$, $\phi_{\text{eff}}(\mbf{r})$, $n(\mbf{r})$, we come
to the system of self--consistent Euler equations, fixing the corresponding
equilibrium distributions.

To simplify the current consideration, let us regard the case of
point dipoles. The molecular density distribution
$\rho^{q}_{\text{mol}}$ in this case is given by
\be
\rho^{q}_{\text{mol}}(\mbf{r})=e\delta\left(\mbf{r}-\frac{\mbf{l}}{2}\right)-
e\delta\left(\mbf{r}+\frac{\mbf{l}}{2}\right)\simeq -e\,
\mbf{l}\mbf{\nabla}\delta(\mbf{r}) =-\mbf{d}\mbf{\nabla}\delta(\mbf{r})
\ee
where $\mbf{d}$ is the dipole moment of the molecule. Then
\be
\phi(\mbf{r},\mbf{d})=-\int d\mbf{r}'\,\phi_{\text{eff}}
(\mbf{r}')\,\mbf{d}\nabla_{\mbf{r}'}\delta(\mbf{r}'-\mbf{r})=
\mbf{d}\nabla\phi_{\text{eff}} (\mbf{r}) \ee this equation together with the
linearized version of \eq{eq:el3} leads to the following expression for the media
charge density $\mu({\bf r})$:
\be \mu(\mbf{r})=\int\int
d\mbf{r}'\,\frac{d\Omega_{\mbf{d}}}{4\pi}\, \beta \;
\mbf{d}\nabla_{\mbf{r}'}\phi_{\text{eff}}(\mbf{r}')
\;\mbf{d}\nabla_{\mbf{r}}\delta(\mbf{r}-\mbf{r}')
=\frac{d^2\beta}{3}\nabla\Big(n(\mbf{r})\nabla \phi_{\text{eff}}
(\mbf{r})\Big) \ee

Inserting the last relation into the Hamiltonian \eq{eq:el_ham} and minimizing the
corresponding expression, we get
\be
\barr{lll} \disp \frac{\delta H}{\delta n} & = & \disp -a\Delta
n(\mbf{r})+\frac{\delta W}{\delta{n}}+ \int_{v_{\text{in}}} d\mbf{r}'\,
\chi^{-1}_{\text{in}}(\mbf{r},\mbf{r}') n(\mbf{r}')-\frac{d^2\beta}{3} \nabla
\varphi(\mbf{r})\nabla \phi_{\text{eff}}(\mbf{r})=0 \medskip \\ \disp \frac{\delta
H}{\delta \phi_{\text{eff}}} & = & \disp \frac{d^2\beta}{3}
\nabla\Big(n(\mbf{r})\nabla \varphi(\mbf{r})\Big)-
\lambda \Delta \phi_{\text{eff}}+\Delta \varphi(\mbf{r})=0 \medskip \\
\disp \frac{\delta H}{\delta{\varphi}} & = & \disp \Delta
\phi_{\text{eff}}+\frac{4\pi}{\varepsilon}\rho^{q}_{\text{solute}}(\mbf{r})+
\frac{d^2\beta}{3} \nabla \Big(n(\mbf{r})\nabla \phi_{\text{eff}}(\mbf{r})\Big)=0
\earr \label{eq:el7}
\ee
This system should be provided with some physically justified boundary conditions,
for example
\be
\frac{\partial \varphi(\mbf{r})}{\partial \mbf{r} } \to 0 \quad \mbox{for
$|\mbf{r}|\to \infty$}; \qquad \phi_{\text{eff}}(\mbf{r}) \to 0 \quad \mbox{for
$|\mbf{r}|\to \infty$}; \qquad n(\mbf{r}) \to \bar{n} \quad \mbox{for $|\mbf{r}|\to
\infty$}
\ee

The numerical computation of the solvation and interaction free energies of charged
extended objects in the dipolar fluctuating media (in the water) based on the set of
derived equations \eq{eq:el7} is in progress and will be the subject of our
forthcoming publication \cite{sineta}.

It is easy to check that in the limit of the weak electrostatic
field equations \eq{eq:el7} lead to the standard electrostatics in
the continuous media with an effective dielectric
density--dependent permittivity. Actually, when the field
$\varphi_{\rm eff}$ is small, we may assume \be \varphi\sim
\phi_{\rm eff} \label{eq:el8} \ee what means that \be \nabla
\varphi\, \nabla\phi_{\rm eff} \sim \phi_{\rm eff}^2
\label{eq:el9} \ee We have omitted in \eq{eq:el8}--\eq{eq:el9} all
numerical prefactors, supposing for simplicity that all quantities
are dimensionless.

Taking into account \eq{eq:el8}--{eq:el9}, we can rewrite the system \eq{eq:el7} in
the lowest order with respect to the electrostatic field:
\be
\barr{r} \disp -a\Delta n(\mbf{r})+\frac{\delta W}{\delta{n}}+ \int_{v_{\text{in}}}
d\mbf{r}'\, \chi^{-1}_{\text{in}}(\mbf{r},\mbf{r}') n(\mbf{r}')=0 \medskip \\
\disp \nabla\Big(\varepsilon(n(\mbf{r}))\nabla \varphi(\mbf{r})\Big)+
\frac{4\pi}{\varepsilon}\rho^{q}_{\text{solute}}(\mbf{r})=0 \earr \label{eq:el10}
\ee
where
\be
\varepsilon(n({\bf r})) \approx 1 + \frac{d^2\beta}{3} n({\bf r}) \label{eq:el12}
\ee
The closed set of equations \eq{eq:el10}--\eq{eq:el12} describes the electrostatics
with the dielectric permittivity linearly proportional to the solvent density.

Let us stress that our derivation implies the absence of the correlations between
the orientational degrees of freedom of water molecules in the bulk and near the
solute surface---only in this case we can pre-average the density $\mu({\bf r})$ in
\eq{eq:el4} over the orientational degrees of freedom. We support our supposition by
the conclusion made in \cite{madan,wallquist} on the basis of extensive numerical
simulations: the rearrangement of the orientational degrees of freedom of water
molecules near the solute surfaces plays the secondary role in the thermodynamics of
solvation (see also \cite{levy} for more discussions).

To conclude with, let us repeat that in the current work the most attention is paid
to the consideration of the pure hydrophobic (cavitation) effect. In particular, the
main achievements are as follows.

1. The method described in the paper provides the constructive basis for the
consistent description of solvation effects of hard solute molecules of any shape.
One sees that the continuous two--length scale fluctuational approach with only two
free parameters tuned to the Monte-Carlo results on solvation of hard spheres,
manages to describe the known cavitation contributions to the solvation free energy
of alkanes without any additional adjustment of these parameters.

2. In the framework of the same approach we have considered two auxiliary problems:
a) the computation of the solvation free energy of solutes of cylindric shape, and
b) the computation of the free energy of interactions of two spheres separated by
some distance. The consideration of the problem (a) allows one to conclude that the
depth of the "skin layer", i.e. the penetration length of the fluctuating field into
the body of the solute is of order of $1.4\xi$ \AA, where $\xi$ is the
characteristic length scale of the theory (the location of the first maximum in the
bulk correlation function of the solvent) while the investigation of the problem (b)
permits us to make conjectures about the potential of the mean force acting between
hard spheres in the fluctuating media.

3. The developed approach can be extended to take into account the dispersion part
of the solute--solvent interactions, as well as the electrostatic contributions of
charged solute molecules. We describe the basic steps towards the implementation of
both these contributions into the current approach.

\begin{acknowledgments}
The authors are very grateful to I. Erukhimovich, I. Bodrenko, O. Khoruzhy, V.
Zosimov for valuable discussions. We also would like to thank M. Levitt and C. Queen
for useful comments and suggestions.
\end{acknowledgments}


\begin{thebibliography}{99}

\bibitem{george}  P. George, C.W. Bock, M. Trachtman, J. Comp. Chem. {\bf 3}, 283 (1982)
\bibitem{gav} A. Gavezzotti, Modelling Simul. Mater. Sci. Eng., {\bf 10}, R1 (2002)
\bibitem{finkelstein} A. Finkelstein, L. Pereyaslavets, in press
\bibitem{surf} T. Ooi, Oobatake, G. Nemethy, H.A. Sheraga, PNAS (USA), {\bf 84}, 3086
(1987)
\bibitem{priv} G. Makhatadze, P.L. Privalov, Adv. Prot.ein Chem., {\bf 47}, 307 (1995)
\bibitem{msa} M.L. Connolly, Science, {\bf 221}, 709 (1983)
\bibitem{karp} T. Lazaridis, M. Karplus, J. Mol. Biol., {\bf 288}, 477 (1999)
\bibitem{rosen} Y. Rosenfeld, J. Chem. Phys. {\bf 98}, 8126 (1993); Y. Rosenfeld, P. Tarazona,
Mol. Phys. {\bf 95}, 141 (1998)
\bibitem{3drism} Q. Du, D. Beglov and B. Roux, J. Phys. Chem. B {\bf 104}, 796 (2000)
\bibitem{chand4} D.M. Huang, Ph.L. Geissler and D. Chandler, J. Phys. Chem. B
{\bf 105}, 6704 (2001)
\bibitem{spt} H.Reiss, H.L. Frisch, J.L. Lebowitz, J. Chem. Phys. 1959, {\bf 31}, 369
\bibitem{spt1} R.A. Pierotti, Chem. Rev., {\bf 76}, 717 (1976)
\bibitem{pratt} G. Hummer, S. Garde, A. Garc\`ia, A. Pohorille, L. Pratt, PNAS
(USA),{\bf 93}, 8951 (1996)
\bibitem{chand1} D. Chandler Phys. Rev. E {\bf 48}, 2898 (1993)
\bibitem{chand2} K. Lum, D. Chandler and J.D. Weeks, J. Phys. Chem. B {\bf 103}, 4570 (1999)
\bibitem{chand3} P.R. Wolde, S.X. Sun and D. Chandler, Phys. Rev. E {\bf 65}, 011201 (2001)
\bibitem{sineta} G. Sitnikov, S.Nechaev, M. Taran, in preparation
\bibitem{ll_1} L.D.Landau, E.M. Lifshitz, {\it Mechanics}, Theoretical Physics, vol.
1, (Boston: Butterworth-Heinemann, 1976)
\bibitem{sarkisov} G.N. Sarkisov, Sov. Phys. Uspekhi, {\bf 169}, 625 (1999)
\bibitem{corr_water} A.K. Soper, Chemical Physics {\bf 258}, 121 (2000)
\bibitem{pearson} R.G. Pearson, J. Am. Chem. Soc. {\bf 108}, 6109 (1986)
\bibitem{py} J.K. Percus, G.J. Yevic, Phys. Rev. {\bf 110}, 1 (1957)
\bibitem{PYanalyt} M.S. Wertheim, Phys. Rev. Letters {\bf 10}, 321 (1963)
\bibitem{PYnum} J. Throop, R.J. Bearman,  J. Chem. Phys. {\bf 42}, 2408 (1964)
\bibitem{irisa} M.Irisa, K. Nagayama, F. Hirata, Chem. Phys. Lett., {\bf 207}, 430
(1993)
\bibitem{pais} A.A.C.C. Pais, A. Sousa, M.E. Eus\'ebio, J.S. Redinha, Phys. Chem.
Chem. Phys., {\bf 3} (2001), 4001
\bibitem{lee} B. Lee, Biopolymers, {\bf 24}, 813 (1985)
\bibitem{widom} B. Widom, Chem. Phys., {\bf 86}, 869 (1982)
\bibitem{levy} E. Gallicchio, M.M Kubo, R.M. Levy, J. Phys. Chem., {\bf 104}, 6271
(2000)
\bibitem{ozrin} V. Ozrin, private communication
\bibitem{halgren} T.A. Halgren, J.
Comput. Chem., {\bf 17} 490; 520; 553; 587; 616 (1996)
\bibitem{dill} N.T. Southall, K.A. Dill, Biophys. Chem. {\bf 101}--{\bf 102}, 295 (2002)
\bibitem{shi} S. Shimizu, H.S. Chan, Proteins: Structure, Function and Genetics,
{\bf 48}, 15 (2002)
\bibitem{ll_5} L.D. Landau, E.M. Lifshitz, {\it Statistical Physics, part I}, Course
in Theoretical Physics, vol. 5, (Boston : Butterworth-Heinemann, 1980)
\bibitem{khkh} A.R. Khokhlov, K.A. Khachaturian, Polymer, {\bf 23}, 1793 (1982)
\bibitem{ll_8} L.D. Landau, E.M. Lifshitz, L.P.Pitaevskii, {\it Electrodynamics of
continuous media}, Course in Theoretical Physics, vol. 8, (Boston: Elsevier, 1984)
\bibitem{madan} B. Madan, B. Lee, Biophys. Chem., {\bf 51}, 279 (1994)
\bibitem{wallquist} S. Durell, A.Wallquist, Biophys. J., {\bf 71}, 1695 (1996)

\end{thebibliography}
\end{document}